\documentstyle[12pt]{article}
\begin{document}
\baselineskip 18pt
\begin{titlepage}
\centerline{\Large\bf $SO(5)$ structure of p-wave superconductivity}
\centerline{\Large\bf for spin-dipole interaction model}
\vspace{2cm}
\centerline{  Hong-Biao Zhang$^{1,2}$, Mo-Lin Ge$^{2}$, and Kang-Xue$^{1}$}
\vspace{0.5cm}
{\small
\centerline{\it 1. Department of Physics, Northeast Normal University,}
\centerline{\it Changchun, Jilin 130024, P.R.China}
\centerline{\it 2. Theoretical Physics Division, Nankai Institute of Mathematics,}
\centerline{\it Nankai University, Tianjin 300071, P.R.China}

}
\vspace{2cm}

\centerline{\bf Abstract}
A closed $SO(5)$ algebraic structure in the the mean-field
form of the Hamiltonian the pure p-wave superconductivity
is found that can help to diagonalized by making use of the
Bogoliubov rotation instead of the Balian-Werthamer approach.
we point out that the eigenstate is nothing but $SO(5)$-coherent
state with fermionic realization. By applying the approach to the
Hamiltonian with dipole interaction of Leggett the consistency
between the diagonalization and gap equation is proved through the
double-time Green function. The relationship between the s-wave
and p-wave superconductivities turns out to be recognized through
Yangian algebra, a new type of infinite-dimensional algebra.

\vspace{1cm}

PACS number(s):  03.65.$-$w,$\;\;$03.65.Fd,$\;\;$74.20.Fg,$\;\;$74.20.$-$z

 Electronic address: hbzhang@eyou.com
\vspace{3cm}

\pagebreak


 The p-wave superconductivitiy theories and their applications to liquid
 $^{3}H_{e}$ have intensively been studied in many initial literatures,
 for example, in Refs\cite{brinkman}-\cite{balian}. As was pointed out
 by Anderson and Brinkman\cite{brinkman} that the BW
 formalism\cite{anderson} underlies all the
 further models in the field. The Hamiltonian takes the Aderson reduced
 form: $H=H_{0}+V$
 \begin{equation}
  \label{H}
  H_{0}=\sum\limits_{{\bf k},{\alpha}} \epsilon_{\bf k} n_{{\bf k}\alpha},
  \;\;\;\;\;\;
  V=\frac{1}{2}\sum\limits_{{\bf k},{\bf k}'{\alpha},{\beta}} V_{{\bf k}{\bf k}'}
  a^+_{{\bf k}'{\alpha}}a^+_{-{\bf k}'{\beta}}
  a_{-{\bf k}{\beta}}a_{{\bf k}{\alpha}}
 \end{equation}
 where $\epsilon_{\bf k}=\frac{{\bf k}^2}{2m}-\mu$,
 $\alpha,\beta=\uparrow,\downarrow$,
 and for p-wave $V_{{\bf k}{\bf k}'}= -3V_{1}(k,k'){\bf n}\cdot{\bf n}'$
 $({\bf n}=\frac{\bf k}{k})$. To explain the p-pair interaction
 Leggett\cite{leggett} introduced an useful algebra, but not closed.
 Meanwhile, the dipole type of interaction was proposed that naturally
 distinguishes the energy difference for ABM and BW phases\cite{leggett}
 -\cite{lee} and gives the correct spin dynamics. All the theories look
 working perfectly, but with growing interest of the applications of
 the currently algebraic method there still desirable efforts could be made
 to yield further understanding. In this paper we would like to
 show the following results.
 a) The set obeying by $S_{i}({\bf k})=a^{+}_{{\bf k}{\alpha}}
                    ({\sigma}_{i})_{\alpha\beta}a_{{\bf k}{\beta}}$
                    and
 $T_{i}({\bf k})=a_{{-\bf k}\alpha}({\sigma}_{2}{\sigma}_{i})_{\alpha \beta}
 a_{{\bf k}\beta}$ as well as their conjugates\cite{leggett} forms $SO(5)$
 algebra ($i=0,1,2,3, \sigma_0={\bf 1}$ and summation over the repeat $\alpha$
 and $\beta$) that is larger than the usual $U(1) \otimes SU^{(L)} \otimes SU^{(S)}$ as
shown in Ref.\cite{volovik}. Equipping with the algebraic structure for
the lower pair excitation we then diagonalize the Hamiltonian eq.(\ref{H})
that together with the invariance $U(1) \otimes SU^{(L)} \otimes SU^{(S)}$
by using the algebraic average method (AAM, general Bogoliubov rotation)
yield all the known results.
 b) We show that the eigenfunction  of eq.(\ref{H}) is
$SO(5)$-coherent state with fermionic realization, hence the coherence
property related to eq.(\ref{H}) origins in the closed $SO(5)$ structure.
 c) The above calculation can be applied to the dipole interaction
Hamiltonian of Leggett. There appears nontrivial consistency between the
diagonalization of microscopic-form of $H_{D}$ by AAM and the gap equation.
d) Finally, in difference from the $SO(5)$ unification of
Zhang et al\cite{zhang}\cite{scalapinp} we attempt to find what price we
have to pay in order to form an assumed unification involving both s-wave
and p-wave superconductivities (as shown in Ref.\cite{leggett}), in stead of
the claimed transition between s-wave superconductivity and AF phase shown
in Refs.\cite{zhang}\cite{scalapinp}.

1) Observing the algebra defined in Ref.\cite{leggett}, if
$T_0({\bf k})$ is picked up, it becomes closed. Defining
 $\bar{\bf S}({\bf k})=\frac{1}{2}[{\bf S}({\bf k})+ {\bf S}(-\bf k)]$,
 $Q{({\bf k})}=\frac{1}{2}(S_0 ({\bf k}) + S_0 ({-\bf k})-2)$ it can be
 checked that the set
($Q({\bf k})$,${\bar{\bf S}}(\bf k)$,${\bf T}(\bf k)$,${\bf T}^{\dag}(\bf k)$)
forms $SO(5)$ algebra:
 $$
 [I_{ab} ({\bf k}), I_{cd} ({\bf k}')]=-i\delta{(\bf k -\bf k')}
 (\delta_{ac} I_{bd} ({\bf k}) + \delta_{bd} I_{ac} ({\bf k})
  -\delta_{ad} I_{bc} ({\bf k}) - \delta_{bc} I_{ad} ({\bf k}) )
 $$
 where $I_{ab}({\bf k})$ = $-I_{ba}({\bf k})$ ($a,b=1,2,3,4,5$)
takes the form:
\begin{equation}
\label{matrix}
\left(
\begin{array}{ccccc}
0&&&&\\
-\frac{1}{2}(T_x^{\dag} ({\bf k}) + T_x ({\bf k})) &0&&&\\
-\frac{1}{2}(T_y^{\dag} ({\bf k}) + T_y ({\bf k})) &
-\bar{S}_z ({\bf k}) &0&&\\
-\frac{1}{2}(T_z^{\dag} ({\bf k}) + T_z ({\bf k})) &
\bar{S}_y ({\bf k}) &-\bar{S}_x ({\bf k}) &0&\\
Q({\bf k})&\frac{1}{2i}(T_x ({\bf k}) - T_x^{\dag} ({\bf k}))&
\frac{1}{2i}(T_y ({\bf k}) - T_y^{\dag} ({\bf k}))&
\frac{1}{2i}(T_z ({\bf k})-T_z^{\dag} ({\bf k}))&0\\
\end{array}
\right)
\end{equation}
The Hamiltonian eq.(\ref{H}) can then be written in the form:
 \begin{equation}
 \label{H1}
 H= \sum\limits_{\bf k} \epsilon_{\bf k}(Q({\bf k})+1)
 +\frac{1}{4}\sum\limits_{{\bf k},{\bf k}'} V_{{\bf k}'{\bf k}}
 {\bf T}^{\dag}({\bf k}) \cdot {\bf T}({\bf k}')
 \end{equation}
Noting that the mean-field approximation is enough to obtain gap equation
since we work at the equilibrium state. Using\cite{solomon}
 $$ AB \simeq A\langle B\rangle+\langle A \rangle B-
 \langle A \rangle \langle B\rangle $$
eq.(\ref{H1}) can be linearized with respect to $SO(5)$ generators
$$
H_{mf}=\sum\limits_{\bf k}\{H({\bf k})-E_{*}({\bf k}) \}
\;\;\;\;\;\;\;\;\;\;\;\;\;\;\;\;\;\;\;\;\;
$$
with
\begin{equation}
\label{Hk}
 H({\bf k})=\epsilon_{\bf k}Q({\bf k})
 +{\bf {\Delta}}({\bf k})\cdot {\bf T}^{\dag}({\bf k})
 + {\bf {\Delta}}^{*}({\bf k}) \cdot {\bf T}({\bf k})
\end{equation}
\begin{equation}
\label{E*}
 E_{*}({\bf k})=\epsilon_{\bf k}
 - {\bf {\Delta}}({\bf k}) \cdot \langle {\bf T}^{\dag}({\bf k}) \rangle
 \;\;\;\;\;\;\;\;\;
\end{equation}
where ${\bf {\Delta}}({\bf k})
 =\frac{1}{4}\sum\limits_{{\bf k}'} V_{{\bf k}'{\bf k}}
 \langle {\bf T} (\bf k) \rangle $
and $\langle {...}\rangle$ represents the average over both quantum states
and thermodynamics. We emphasized that the set $\Lambda=
\{ \frac{i}{\sqrt{2}}T_{3}({\bf k}),\frac{-i}{\sqrt{2}}T^{\dag}_{3}({\bf k}),
-Q({\bf k}) \}$ i.e. $\{ -i\sqrt{2}\pi_{z},i\sqrt{2}\pi^{\dag}_{z}, -Q \}$
in Refs.\cite{zhang}\cite{scalapinp} forms the quasi-spin $\Lambda$ for pairs.
$\Lambda$ does not commute with spin operators ${\bf S}({\bf k})$ that
give rise to $T^{\dag}_{\pm}({\bf k})$ and $T_{\pm}({\bf k})$ which are beyond
two $SU(2)$'s and the total set forms $SO(5)$.
In order to make the diagonalization of eq.(\ref{Hk}) we introduce
the unitary transformation such that $W^{\dag}(\xi_{\bf k})H({\bf k})W(\xi_{\bf k})$
becomes diagonal for any ${\bf k}$. Following the general strategy\cite{feng}
we introduce the $SO(5)$-coherent operator:
\begin{equation}
\label{xik}
 W(\xi_{\bf k})=\exp\{\xi_{\bf k}[{{\bf d}(\bf n)}\cdot {\bf T}^{\dag}
 ({\bf k})] -h.c.\}
 \end{equation}
  where $ \xi_{\bf k}$ = $r_{\bf k}e^{i\lambda_{\bf k}}$,
 ${\bf d}({\bf n})=(\sin{\Theta_{\bf k}}\cos{\Phi_{\bf k}},
 \sin{\Theta_{\bf k}}\sin{\Phi_{\bf k}},\cos{\Theta_{\bf k}})$,
 $\Theta_{\bf k}$ and $\Phi_{\bf k}$ are angulars in spin space for
 a given momentum ${\bf k}$. $\lambda_{\bf k}$ is a parameter to be
 determined by the gap equation.
 Taking the commutation relations for $SO(5)$ into account after lengthy
 but elementary calculations we derive
 \begin{equation}
 \label{w}
 W(\xi_{\bf k})^{-1}H({\bf k})W(\xi_{\bf k})=-E_{\bf k}Q(\bf k),
\;\;\;\;\;
 E_{\bf k}=\sqrt{{\epsilon}^{2}_{\bf k}+|{\bf {\Delta}}({\bf k})|^{2}}
\end{equation}
where
\begin{equation}
\label{tan}
\tan{2r_{\bf k}}=\frac{|{\bf {\Delta}}({\bf k})|}
{{\epsilon}_{\bf k}},
\;\;\;\;\;
{\bf {\Delta}}({\bf k})
=\frac{1}{4}\sum\limits_{{\bf k}'} V_{{\bf k}'{\bf k}}
 \langle {\bf T} (\bf k) \rangle
=-\frac{1}{2}|{\bf {\Delta}}({\bf k})|
e^{i\lambda_{\bf k}}{\bf d}(\bf n)
\end{equation}
The eigenstate is given by
\begin{equation}
\label{xi}
 |\xi\rangle
 =\otimes_{\bf k} |\xi_{\bf k} \rangle ,
\;\;\;\;\;
|\xi_{\bf k} \rangle=W(\xi_{\bf k})|{\rm vac} \rangle
\end{equation}
At temperature $T=0$, the vacuum state is $|{\rm vac} \rangle = |0,0
\rangle_{\bf k} \equiv |n_{{\bf k}\alpha}=0,n_{{-\bf k}\alpha}=0 \rangle$.
The expectation value
$\langle{\bf T}({\bf k})\rangle=\langle\xi_{\bf k}|{\bf T}({\bf k})|\xi_{\bf k}\rangle
=\sin2r_{\bf k} e^{i\lambda_{\bf k}}{\bf d}(\bf n)$ yields the well-known
gap equation at $T=0$:
\begin{equation}
\label{gap0}
{\bf {\Delta}}({\bf k})
=-\sum\limits_{\bf k'}V_{\bf k' \bf k}
\frac{\bf {\Delta}({\bf k}')}{2E_{{\bf k}'}}
\end{equation}
To satisfy eq.(\ref{gap0}) we simply choose $\lambda_{\bf k}=\lambda=$constant
henceforth, for finite temperature, making use of the double-time
Green function we obtain
$\langle n_{{\bf k}\alpha}\rangle = \frac{1}{2}[1-\frac{\epsilon_{\bf k}}
{E_{\bf k}}\tanh(\frac{1}{2}\beta E_{\bf k})]$ and
$\langle{\bf T}({\bf k})\rangle=\langle\xi_{\bf k}|{\bf T}({\bf k})|\xi_{\bf k}\rangle
=\sin2r_{\bf k} e^{i\lambda} \tanh(\frac{1}{2}\beta E_{\bf k}){\bf d}(\bf n)$
where $\beta=\frac{1}{kT}$ and $k$ is the Boltzmann constant.
Therefore the gap equation reads
\begin{equation}
\label{gap1}
{\bf {\Delta}}({\bf k})=-\sum\limits_{{\bf k}'} V_{{\bf k}{\bf k}'}
\frac{\bf {\Delta}({\bf k}')}{2E_{{\bf k}'}}
\tanh(\frac{1}{2}\beta E_{{\bf k}'})
\end{equation}
The $SO(5)$ coherent state $| \xi_{\bf k} \rangle$ eq.(\ref{xi}) gives
$$
 |\xi_{\bf k} \rangle=W(\xi_{\bf k})|0,0 \rangle
     =\cos^{2}r_{\bf k}|0,0 \rangle - e^{i2\lambda}\sin^{2}r_{\bf k}
 |\uparrow\downarrow,\uparrow\downarrow \rangle
 \;\;\;\;\;\;\;\;\;\;\;\;\;\;\;
 $$
 \begin{equation}
 \label{CS}
 +\frac{i}{2}e^{i\lambda}
 \sin{2r_{\bf k}}\{\cos{\Theta_{\bf k}}(|\uparrow,\downarrow \rangle
 +|\downarrow,\uparrow \rangle)-\sin{\Theta_{\bf k}}e^{-i\Phi_{\bf k}}
 |\uparrow,\uparrow \rangle+\sin{\Theta_{\bf k}}e^{i\Phi_{\bf k}}
 |\downarrow,\downarrow \rangle \}
\end{equation}
Let us distinguish two cases: In the BW phase, to satisfy the gap equation
it holds ${\bf d}(\bf n)$=${\bf n}$, $\Theta_{\bf k}=\theta_{\bf k}$ and
$\Phi_{\bf k}={\psi}_{\bf k}$ that correspond intuitively to Cooper pair
with total angular momentum $J=0$ and has an isotropic
gap, $|{\bf {\Delta}}({\bf k})|e^{i\lambda}$=c-number.
The wave function of the BW solution in the conventional notation reads
$$
 |\xi_{\bf k} \rangle=\frac{E_{\bf k}+\epsilon_{\bf k}}
 {2E_{\bf k}}|0,0 \rangle - e^{i2\lambda}
 \frac{E_{\bf k}-\epsilon_{\bf k}}{2E_{\bf k}}
 |\uparrow\downarrow,\uparrow\downarrow \rangle  \;\;\;\;\;\;\;\;\;\;
$$
\begin{equation}
\label{bw}
 -e^{i\lambda}\frac{i|{\bf {\Delta}}(\bf k)|}{2E_{\bf k}}
 \sqrt{\frac{8\pi}{3}}
 \{Y_{11}|\downarrow,\downarrow \rangle
 -\frac{1}{\sqrt{2}}Y_{10}(|\uparrow,\downarrow \rangle
 +|\downarrow,\uparrow \rangle)
 +Y_{1-1}|\uparrow,\uparrow \rangle \}
\end{equation}
In the AM case, there is another solution of the gap equation
by taking
$|{\bf {\Delta}}(\bf k)|e^{i(\lambda+\frac{\pi}{2})}=Y_{11}$
and $\sin{\Theta_{\bf k}}=0$, it is non-ESP state:
$$
 |\xi_{\bf k} \rangle=\frac{E_{\bf k}+\epsilon_{\bf k}}
 {2E_{\bf k}}|0,0 \rangle + e^{i2\lambda}
 \frac{E_{\bf k}-\epsilon_{\bf k}}{2E_{\bf k}}
 |\uparrow\downarrow,\uparrow\downarrow \rangle
$$
\begin{equation}
\label{esp}
 + \frac{Y_{11}}{2E_{\bf k}}(|\uparrow,\downarrow \rangle
 +|\downarrow,\uparrow \rangle)
\end{equation}

However, the solution for $\cos{\Theta_{\bf k}}=0$ appears only under an
applied magnetic field. For instance, when ${\bf B}=\mu B {\bf e}_{z}$,
the Hamiltonian becomes $H_{B}=\sum\limits_{\bf k} H({\bf k})-
\mu B\sum\limits_{\bf k}(a^{+}_{{\bf k}\uparrow}a_{{\bf k}\uparrow}
-a^{+}_{{\bf k}\downarrow}a_{{\bf k}\downarrow})$ and
$W^{\dag}H_{B}W=\frac{1}{2}E_{\bf k}(1+\frac{\mu B}{\epsilon_{\bf k}})
(n_{{\bf k}\downarrow}+n_{{-\bf k}\downarrow})
+ \frac{1}{2}E_{\bf k}(1-\frac{\mu B}{\epsilon_{\bf k}})
(n_{{\bf k}\uparrow}+n_{{-\bf k}\uparrow})-E_{\bf k}$. Through the
double-time Green-function the non-vanishing components of
${\bf {\Delta}}({\bf k})$
are  $\Delta_{\uparrow\uparrow}({\bf k})
 =\frac{1}{4}\sum\limits_{{\bf k}'} V_{{\bf k}{\bf k}'}
  \frac{|\bf {\Delta}({\bf k}')|}{2E_{{\bf k}'}}
  e^{i\lambda} e^{i\Phi_{{\bf k}'}}
  \tanh[\frac{1}{2}\beta E_{{\bf k}'}(1-\frac{\mu B}{\epsilon_{{\bf k}'}})]$ and
$\Delta_{\downarrow\downarrow}({\bf k})$ obtained by changing in
$\Delta_{\uparrow\uparrow}({\bf k})$ by $\pi+\lambda$,
$-\Phi_{{\bf k}'}$
and $-B$, then times (-1). The eigenstate is $\sim Y_{11}$
($\mid {\uparrow\uparrow}\rangle -\mid {\downarrow\downarrow}\rangle$).
If we study s-superconductivity by AAM, then coherence comes from the
$SU(2)$ coherent operators in Ref.\cite{feng}. The AAM is exactly the
usual Bogoliubov transformation.

2) The dipole interaction was proposed to describe the spin dynamics for
liquid $^{3} H_{e}$. The computation in Ref.\cite{leggett} is based on
$\langle T({\bf k}) \rangle =\sum\limits_{i=1}^3 T^{\alpha}_{i}n_{i}$.
However, in the microscopic-form of dipole interaction Hamiltonian
\begin{equation}
\label{HD}
  H_{D}=\frac{2\pi\gamma^{2}}{3}\sum\limits_{\bf k \bf k}V_{\bf k \bf k'}
  ({\bf T}^{\dag}(\bf k) \cdot {\bf T}(\bf k')
 -3{\hat{\bf q}}\cdot {\bf T}^{\dag}(\bf k)
 {\hat {\bf q}}\cdot {\bf T}(\bf k') )
 \end{equation}
where ${\hat {\bf q}}$ is a unit vector along ${\bf n}-{\bf n'}$.
The average formula shown in Ref.\cite{leggett} can no longer be used,
since the operator cannot be expended by $n_{i}$. However, the $SO(5)$
AAM procedure works for eq.(\ref{HD}) with the redefinition
$\Delta_{i}=\sum\limits_{j=1}^{3}(\delta_{ij}
-3{\hat q}_{i}{\hat q}_{j})\langle T_{j}\rangle$. Repeating the process in
1) we find that the ${\bf {\Delta}}({\bf k})$ satisfies the same gap equation
eq.(\ref{gap1}) and the eigenstates take the same form for given
${\bf {\Delta}}({\bf k})$ satisfying eq.(\ref{gap1}), i.e., eq.(\ref{HD})
can be diagonalized with the following relation instead of eq(\ref{tan}):
$$
\left(
\begin{array}{c}
\Delta_{\uparrow\uparrow} ({\bf k}) \\
\sqrt{2}\Delta_{\uparrow\downarrow} ({\bf k}) \\
\Delta_{\downarrow\downarrow} ({\bf k}) \\
\end{array}
\right)
 = \frac{4\pi\gamma^{2}}{3}\sum\limits_{{\bf k}'} V_{{\bf k}'{\bf k}}
 \{ \frac{1}{2}I
$$
\begin{equation}
\label{Delta}
 +\frac{3}{2}D^{j=1}(\alpha=\psi_{{\bf k}{\bf k}'},
 \beta=2\omega_{{\bf k}{\bf k}'},
 \gamma=\pi-\psi_{{\bf k}{\bf k}'}) \}
 \left(
\begin{array}{c}
\langle \frac{1}{2i}T_- ({\bf k}') \rangle \\
\langle \frac{i}{\sqrt{2}}T_z ({\bf k}') \rangle \\
\langle \frac{i}{2}T_+ ({\bf k}') \rangle \\
\end{array}
\right
)
\end{equation}
where the $D^{j=1} (\alpha,\beta,\gamma)$ is the Wigner rotation
function with
the Euler angles $\alpha$, $\beta$ and $\gamma$ and $\hat {\bf q}
=q(\sin{\omega_{{\bf k}{\bf k}'}}\cos{\psi_{{\bf k}{\bf k}'}},
\sin{\omega_{{\bf k}{\bf k}'}\sin{\psi_{{\bf k}{\bf k}'}},
\cos{\omega_{{\bf k}{\bf k}'}}})$. The relations for
$\omega_{{\bf k}{\bf k}'}$, $\psi_{{\bf k}{\bf k}'}$ and
$\alpha$, $\beta$, $\gamma$ have been indicated in eq.(\ref{Delta}).
The $H_{D}$ works well in spin dynamics, but the consistency condition
for the diagonalization of $H_{D}$ and gap equation, in our knowledge,
have not been proved before. Now the eq.(\ref{tan}) is replaced by
eq.(\ref{Delta}) for the Hamiltonian eq.(\ref{HD}). It means that
all the discussion in 1) can be transplanted literally for
${\bf {\Delta}}({\bf k})$.

3) It seems that the $T_{\pm}$ (or $\sim \pi_{\pm}$
in Refs.\cite{zhang}\cite{scalapinp}) in $SO(5)$ may give rise to the
transition between superconductivity and AF state based on the argument
in Refs.\cite{zhang}\cite{scalapinp}. However, it is not the case in the
present model. This is not only because there is not $SO(5)$ invariance
for $H$ or $H_{D}$, but also for deeper reason. Observing the gap equation
for $V_{{\bf k}{\bf k}'} \sim P_{0}$ (constant) and $V_{{\bf k}{\bf k}'}
\sim P_1\sim {\bf n}\cdot{\bf n'}$,
the corresponding wave function $\Psi_0 \sim Y_{00} \chi_{00}$
and $\Psi_1 \sim$ eq.(\ref{bw}) where $\chi_{00}$ is spin singlet.
In our case, the generators of $SO(5)$ work only within p-supeconductivity,
i.e. nothing with s-superconductivity. If we assume there is transition
between $\Psi_{0}$ and $\Psi_{1}$, i.e. the form of gap equation is preserved,
but with the different potentials $ \sim P_{0}$ and $P_{1}$, the relationship occurs
between two states:
\begin{equation}
\label{p}
\Psi_{0} \sim \frac{Y_{00}}{\sqrt{2}}
\left(
\begin{array}{lr}
0&1\\
-1&0\\
\end{array}
\right
) ,
\;\;\;\;
\Psi_1 \sim \frac{1}{\sqrt{3}}
\left(
\begin{array}{lr}
Y_{1-1}&\frac{-1}{\sqrt{2}}Y_{10}\\
\frac{-1}{\sqrt{2}}Y_{10}&Y_{11}
\end{array}
\right
)
=\frac{1}{\sqrt{8\pi}}
\left(
\begin{array}{cc}
\hat {k}_{-}&-\hat {k}_z\\
-\hat{k}_z&-\hat{k}_{+}\\
\end{array}
\right
)
\end{equation}
The connection should be beyond Lie algebra. We find that such a "transition"
may be performed through Yangian\cite{drinfeld1}-\cite{drinfeld3}. Actually,
$$
\hat{k}_{\pm}J_{\mp}\Psi_0 =(\mu_{2}-\mu_{1}+\frac{h}{2})
Y_{1\pm1}\chi_{1\mp1}
$$
\begin{equation}
\label{transition}
\hat{k}_z J_{z}\Psi_{0}=-\frac{1}{2}(\mu_{2}-\mu_{1}+\frac{h}{2})
 Y_{10}\chi_{10}
\end{equation}
where $J_{\alpha}=\mu_{1}S_{\alpha} \otimes 1 +\mu_{2} 1 \otimes S_{\alpha}
-\frac{ih}{4}\epsilon_{\alpha\beta\gamma}(S_{\beta} \otimes S_{\gamma}
-S_{\gamma} \otimes S_{\beta})$ $(\alpha,\beta,\gamma=1,2,3)$ and $S_{\alpha}$ the
spin operators. $\mu_1 $ and $\mu_2$ are arbitrary constants allowed by
Yangian representation theory and play the crucial role in the representation
of Yangian\cite{chari}.
$[S_{\alpha},J_{\beta}]
=i\epsilon_{\alpha\beta\gamma}J_{\gamma}$ and $J_{\gamma}$'s obey the nonlinear
commutation relations\cite{drinfeld1}\cite{ge}. The set $\{S_{\alpha},J_{\beta}\}$
forms Yangian associated with $SU(2)$ denoted by $Y(SU(2))$.
Noting that $J_{\alpha}$ act on quantum
tensor space only. If the set $\{{\bf S},{\bf J}\}$ satisfies Yangian,
so does ${\bf J} + \eta {\bf S}$ with
arbitrary constant $\eta$ that is called translation of Yangian.
By taking an appropriate translation constant we have
\begin{equation}
\label{wave1}
  (\hat{\bf k} \cdot {\bf J}) \Psi_0 = \frac{\sqrt{3}}{2}
  (\mu_{2}-\mu_{1}+\frac{h}{2})\Psi_1 ,
  \;\;\;\;
  (\hat{\bf k} \cdot {\bf J}) \Psi_1 =0 .
\end{equation}
Yangian is an infinte algebra. Therefore any attempt to unify the
superconductivity with different $l$-waves should be through infinite
algebra. For the simply physical realization of $Y(SU(2))$, see Ref.\cite{ge}.

4) In conclusion we believe that the AAM provides an useful approach to discuss
physics concerning pair-particles, especially for the nature of coherence and
consistency between the diagonalization of given Hamiltonian and gap
equation through double-time Green function. Further, this algebraic method
may be extended to Yangian algebra that is natural to describe the transition
between different condensates.

 \vspace{3mm}

This work is in part supported by the National Natural Science Foundation
of China.


\end{titlepage}
\end{document}